\documentclass{article}

\usepackage{times,amsmath,amsfonts,amssymb,latexsym,textcomp}
\usepackage{color}
\usepackage{graphicx}
\usepackage{multirow}
\usepackage{bbm}
\usepackage{tabularx}
\usepackage{amsthm}
\usepackage{dsfont}
\usepackage{float}
\usepackage{array}
\usepackage{placeins}
\usepackage{MnSymbol}
\usepackage[electronic]{ifsym}
\usepackage[T1]{fontenc}
\usepackage[latin9]{inputenc}
\usepackage[english]{babel}

\begin{document}

\title{Locally Causal and Deterministic Interpretations of Quantum Mechanics: Parallel Lives and Cosmic Inflation}
\author{Mordecai Waegell}

\maketitle

\textbf{Abstract:}
Several locally deterministic interpretations of quantum mechanics are presented and reviewed.  The fundamental differences between these interpretations are made transparent by explicitly showing what information is carried locally by each physical system in an idealized experimental test of Bell's theorem.  This also shows how each of these models can be locally causal and deterministic.  First, a model is presented which avoids Bell's arguments through the assumption that space-time inflated from an initial singularity, which encapsulates the entire past light cone of every event in the universe.  From this assumption, it is shown how quantum mechanics can produce locally consistent reality by choosing one of many possible futures at the time of the singularity.  Secondly, we review and expand the Parallel Lives interpretation of Brassard and Raymond-Robichaud, which maintains local causality and determinism by abandoning the strictest notion of realism.  Finally, the two ideas are combined, resulting in a parallel lives model in which lives branch apart earlier, under the assumption of a single unified interaction history.  The physical content of weak values within each model is discussed, along with related philosophical issues concerning free will.

\section{Introduction}

It was once considered a great triumph of physics that nature could be described by a set of local and deterministic rules, a so-called `clockwork' universe in which the entire future state of all physical systems could be exactly predicted with perfect knowledge of their present state.  Physical models of this type were confounding to many theologians and philosophers because they imply that as a physical system, a human being lacks true free will, and that the experience of free will is merely an illusion enjoyed by automata, who are truly only capable of fulfilling their predetermined destiny.  Despite these objections, there was a time when the general consensus among scientists was that the universe was just such a clockwork machine - after all, this description is elegant, and its predictive power was its greatest success.

With the advent of quantum mechanics and its associated uncertainty, doubt was cast on the idea of perfect deterministic predictions.  This restored the possibility that the physics of nature could be consistent with genuine free will, and in the intervening years, the consensus opinion has undergone a paradigm shift away from determinism and back towards free will.  There was no physical grounds for this shift - it is really just a matter of personal preference.  People say things to the effect that deterministic models are `disgusting' or `ridiculous,' or that they would mean an end to physics as field of study, but as T'Hooft so elegantly pointed out, `disgusting' and `ridiculous' are not exactly rigorous mathematical arguments \cite{hooft2013fate}.

The work of John S. Bell \cite{Bell1, Bell2, EPR} put the final nail in the coffin of local realism, but there are many possible explanation of this result - some which preserve Bell's notion of realism at the cost of locality, and others which preserve locality at the cost of realism.

For example, the Bohmian interpretation of quantum mechanics \cite{bohm1952suggested1, bohm1952suggested2} maintains physical realism of all properties of a system by invoking an explicitly nonlocal interaction mechanism called the {\it quantum potential}.  It is important to note that while the Bohm model is nonlocal, it remains a deterministic theory, where one simply `turns the crank' to compute the future evolution of systems.

The subject matter of this article is primarily focused on another class of explanations, in which local causality and determinism are preserved, but Bell's original notion of realism is replaced by a parallel lives ontology for physical systems.  We will also consider a scenario in which Bell's argument is simply nullified.  In all of these pictures, determinism is preserved, and free will is a subjective illusion.

We will also consider the physical interpretation of weak values \cite{aharonov1988result, aharonov1991complete, dressel2015weak} in each of these locally causal and deterministic models.  None of these models include a post-selected vector that propagates from future to past, but nevertheless the information needed to define a weak value is available locally in some of these models.

We begin with a review of Bell's theorem and a corresponding experimental setup to test it.  We explore each subsequent interpretation in detail using this experimental setup as the common frame of reference.

\section{The Bell-GHZ Theorem}
\begin{figure}[h!]
\centering
\caption{The Peres-Mermin Square}
\includegraphics[width=2.5in]{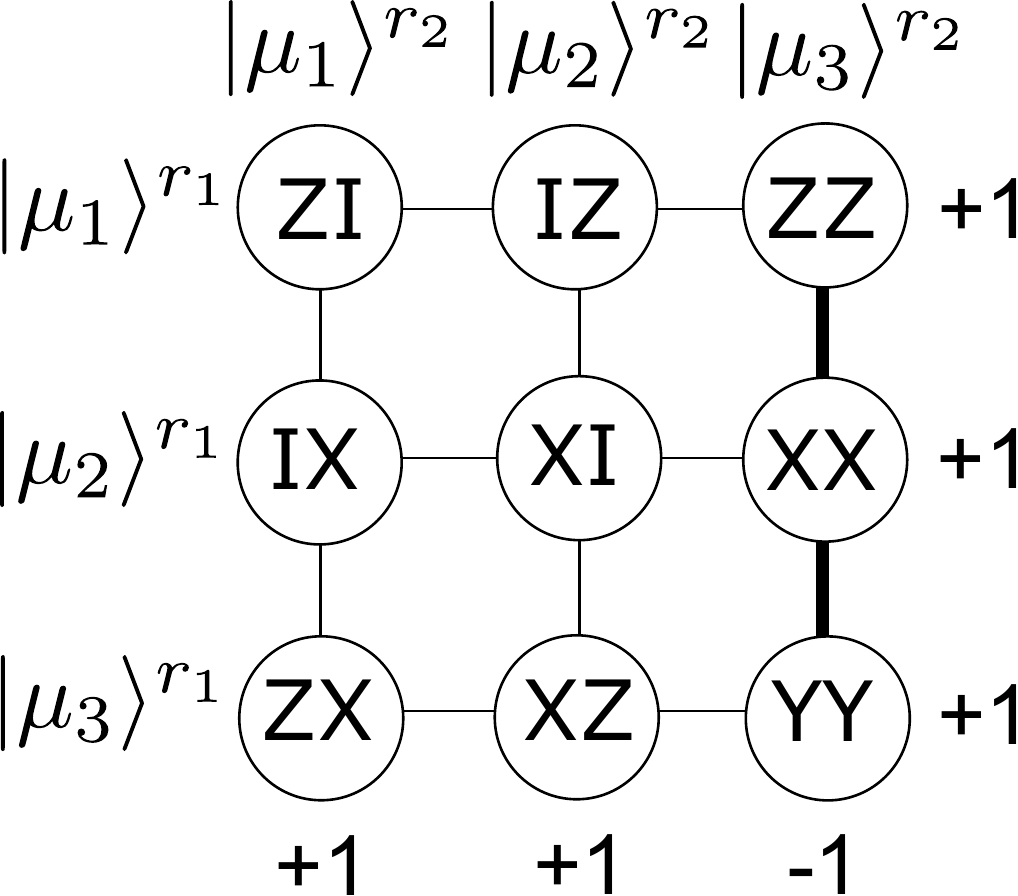}
\label{Square}
\end{figure}
\begin{figure}[h!]
\centering
\caption{Causality diagram for an ideal experimental test of Bell's theorem.  Six physical system that carry local causal signals are shown.  The systems s$_1$ and s$_2$ carry signals from S, r$_1$ and r$_2$ carry random numbers selected by R$_1$ and R$_2$ respectively, and m$_1$ and m$_2$ carry information about the outcome of the measurements at M$_1$ and M$_2$ respectively.  The $\gamma_?$ indicates the possibility that even after it is triggered, the experiment may not be completed as intended if there is some unforeseen disaster.}
\includegraphics[width=3.5in]{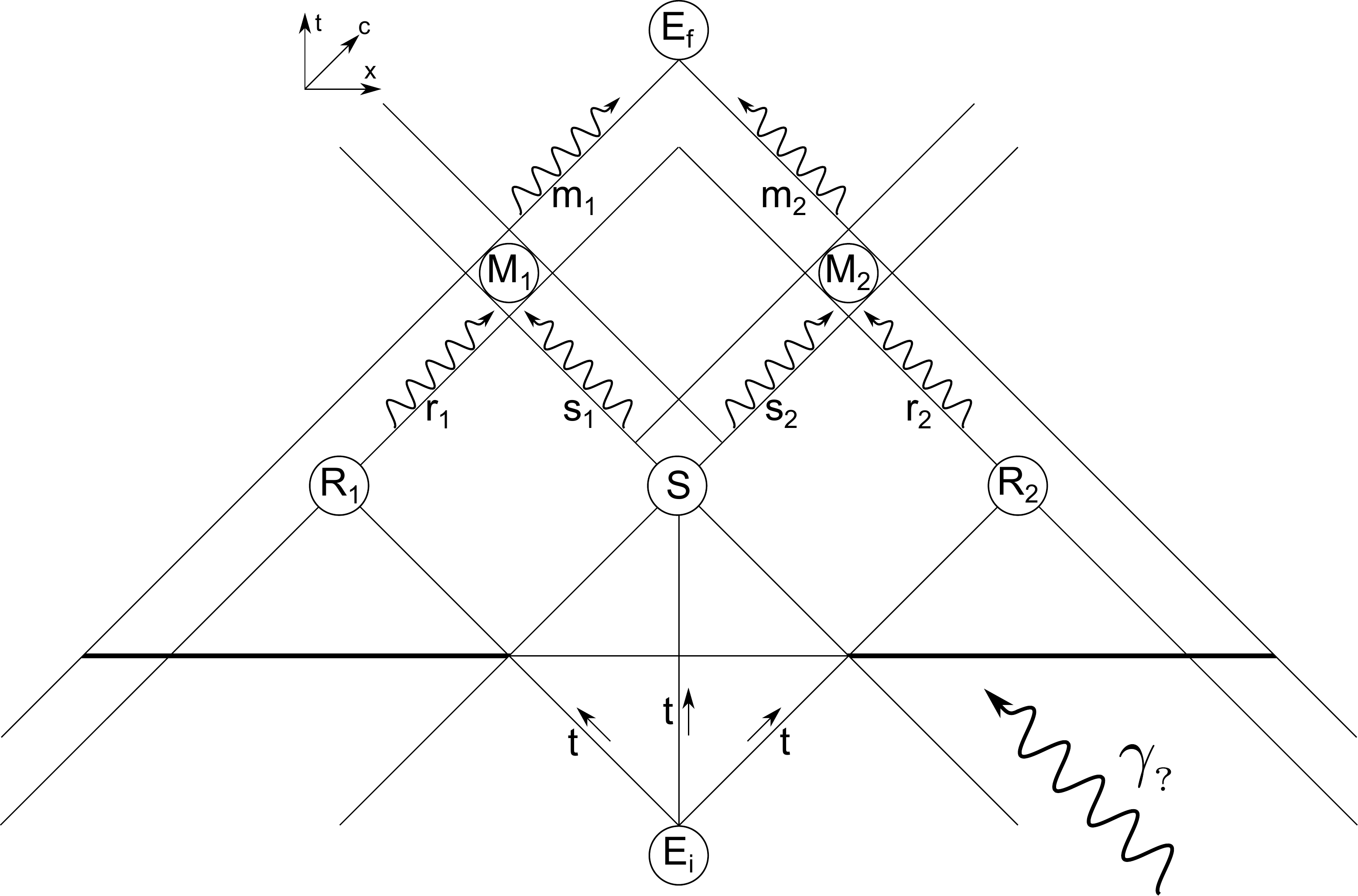}
\label{Causality}
\end{figure}
  \begin{figure}
\centering
\resizebox{\textwidth}{!}{
\begin{tabular}{c|ll}
  \hline$r_1$ & $|\mu_i\rangle^{r_1}_,$  & any \textbf{one} value of $i \in \{1,2,3\}$.  \\
   $r_2$ & $|\mu_j\rangle^{r_2}_,$ &  any \textbf{one} value of $j \in \{1,2,3\}$.   \\
   $s_1$ & $|\psi_1(\{\lambda_n\})\rangle^{s_1}_,$ &  $n=1,\ldots,9$, predicts the outcome for all nine observables of the PMS. \\
   $s_2$ & $|\psi_2(\{\lambda_n\})\rangle^{s_2}_,$ &  the same set $\{\lambda_n\}$ as above. \\
   $m_1$ & $|\phi_1\rangle^{m_1}_,$ &  the outcome predicted by $\{\lambda_n\}$ given $|\mu_i\rangle^{r_1}$. \\
   $m_2$ & $|\phi_2\rangle^{m_2}_,$ &  the outcome predicted by $\{\lambda_n\}$ given $|\mu_j\rangle^{r_2}$. \\
   $E$   & $|\Omega\rangle^E_,$ &\\
\hline
\end{tabular}}
\caption{Information carried by different physical systems in Bell's local realist model is shown in parentheses.  It is impossible to choose a set of nine eigenvalues $\{\lambda_n\}$ such that all six predicted outcomes obey the constraint of (2), which shows that the hidden variables $\{\lambda_n\}$ cannot exist, and this model fails, proving Bell's theorem.}
\label{Bell}
\end{figure}
Bell's Theorem states that local realism is incompatible with quantum mechanics.  The simplest proof of the Bell-GHZ theorem \cite{aravind2002bell, Bell1, GHZ, GHSZ} for two parties, Alice and Bob, makes use of the Peres-Mermin Square (PMS) \cite{Peres24, Mermin_SquareStar, WA_24Rays} of Figure \ref{Square}, which has the following properties:\newline
  (1) - Each row and column of the square represents a possible measurement on two qubits using Pauli observables.\newline
  (2) - The result of the measurement is an eigenvalue (+1 or -1) for each of the three observables in that row or column, with the added constraint that the product of all three values will always be fixed as shown for each row or column.\newline

Figure \ref{Causality} shows a causality diagram of an ideal Bell experiment with two parties, Alice and Bob, operating detectors M$_1$ and M$_2$ respectively.  In this case, the source S prepares four qubits in the state,
\begin{equation}
|\psi\rangle = \left(|0^1\rangle |0^3\rangle + |1^1\rangle |1^3\rangle\right)\left(|0^2\rangle |0^4\rangle + |1^2\rangle |1^4\rangle\right) / 2.
\end{equation}

Qubits 1 and 2 (shown by superscripts) are sent to detector M$_1$, forming the locally propagating signal s$_1$ , while qubits 3 and 4 are sent to detector M$_2$, forming signal s$_2$, such that one member of each correlated Bell pair is sent to either detector.
   Based on a signal from quantum random number generator R$_1$, M$_1$ measures one of the three {\it rows} of the PMS, and based on a signal from R$_2$, M$_2$ measures one of the three {\it columns} of the PMS, such that one observable of the square is always measured by both M$_1$ and M$_2$ at random.
   Here is the crucial idea of the Bell-GHZ theorem:\newline
  (3) - Because of the entanglement correlation of the Bell states, both M$_1$ and M$_2$ always obtain the same outcome for the observable they measure in common.\newline
  (4) - The only explanation for this perfect correlation that obeys local realism is that the outcome for M$_1$ and M$_2$ were predetermined at S, and this information was carried by both s$_1$ and s$_2$.  This information, often called a hidden variable, is simply a list of nine eigenvalues, $\{\lambda_n\}$, corresponding to predictions for the nine observables.\newline
  (5) - It is impossible to choose the list of nine eigenvalue predictions of (4) that will obey property (2), which shows that the hidden variable cannot exist, and thus the experiment does not obey local realism.\newline
  (6) - To see why (2) and (4) contradict each other, consider the product of predicted values for all three rows and all three columns.  According to (2), this product is -1, since it is just the product of the six signs.  According to (4), this product is +1, since all nine eigenvalues will be squared.\newline

  Figure \ref{Bell} shows explicitly what information must be carried by each system in Bell's local realist ontology.

\section{Local Causality and Determinism Assuming Cosmic Inflation}
Now we consider the first of several alternative explanations of how the correlations in the Bell experiment can be explained by local determinism.  The first such model we discuss also preserves realism of the type defined by Bell, but still violates his inequalities.

\begin{figure}[h!]
\centering
\caption{Causality diagram for an ideal experimental test of Bell's theorem with inflation from a singularity.  In this model, the presence or absence of $\gamma_?$ is predetermined, and can be included in the process of defining hidden variables - it is now just part of the list of possible futures that is available locally because $|\Psi_0\rangle$ the entire past light cone of all points in space-time. }
\includegraphics[width=3.5in]{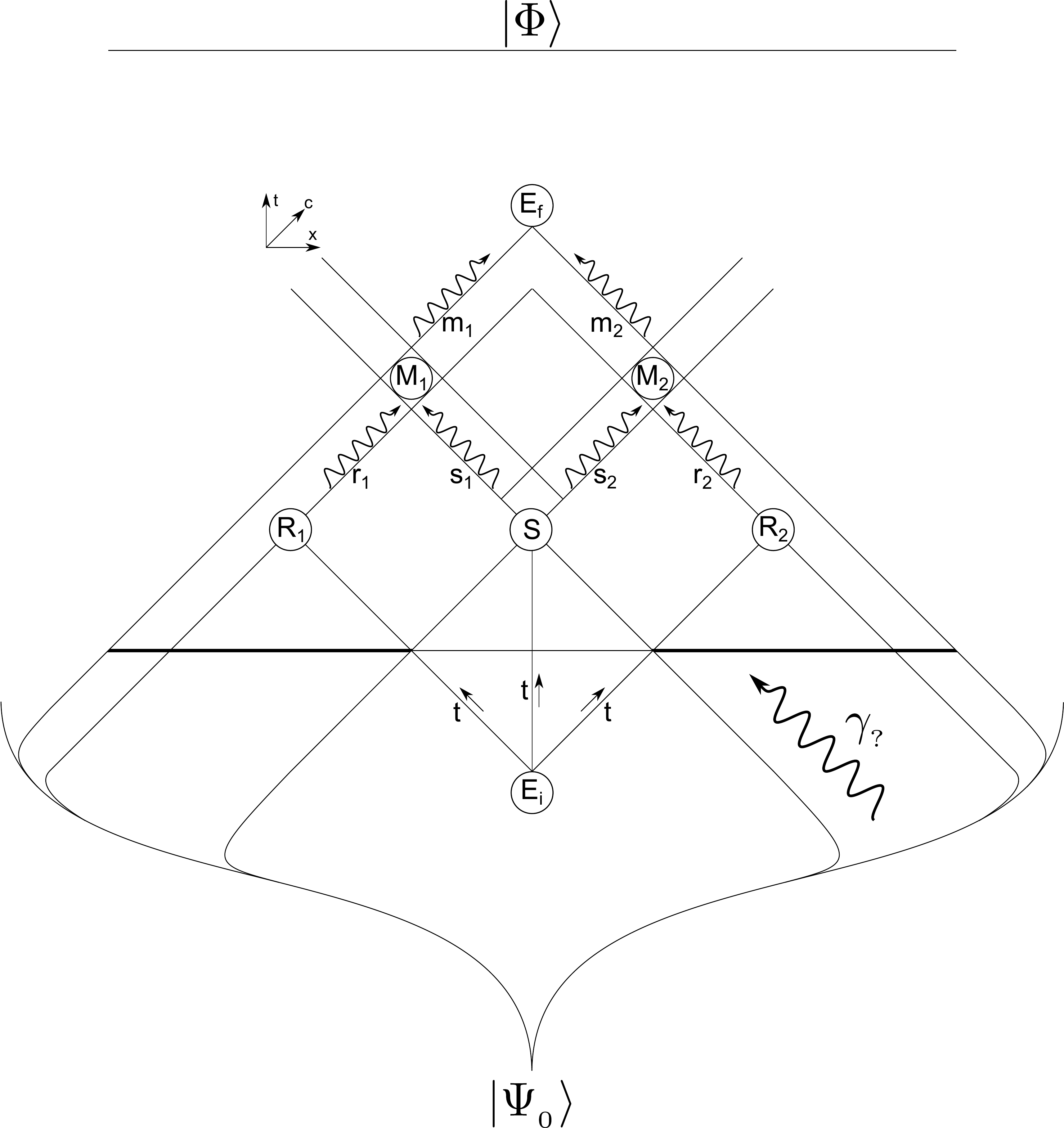}
\label{CausalityInf}
\end{figure}

   The explanation is fairly straightforward.  If we assume that the universe began as a singularity of space, and then underwent cosmic inflation such that points in space stretched apart much faster than $c$, then the initial state of the universe is within the past light-cone of every point in space-time, and it is in fact, the {\it entire} past light cone of every event, as depicted in Figure \ref{CausalityInf}.  This idea has been mentioned by Tipler \cite{tipler2012nonlocality}.

   Under the assumptions of local causality and determinism, the information contained in the initial state amounts to a complete list of all possible future states of the universe, and the probabilities for each to be realized is given by the usual quantum formalism.  We then show explicitly how a locally deterministic mechanism can produce all of the possibilities predicted by quantum theory.  This mechanism does not include wavefunction collapse or a classical limit.  There is a single universal wavefunction that contains a superposition state of all systems.

   We must also inherently assume that cosmic inflation itself is caused by an unknown aspect of the same deterministic mechanism.

   In this model, at the time of the singularity, nature performed a locally causal and deterministic {\it simulation} to determine all possible future states of the universe and their corresponding quantum probability, and then it randomly chose one to be realized.  This choice specifies the final state of the universe and all intervening states as well, and might be considered the one and only collapse in this interpretation.

   This explanation of realism is often called superdeterminism, and cosmic inflation provides a means for this superdeterminism to derive from a locally causal and deterministic mechanism.  In this model, Bell's theorem is effectively nullified because there are no free choices of measurement settings in a superdeterministic realist ontology.

   The following example makes clear how this mechanism can be deterministic, and built only upon information carried by physical systems that propagate obeying local causality.

   Consider the Bell experiment once more.  The mechanism of this model uses complete knowledge of the initial state of the universe to locally compute the list of possible future events, such that suitable hidden variables can be specified for each possibility, and then carried by locally causal signals.  For this experiment, supposing it all goes as the experimenter intends (the lab is not destroyed by $\gamma_?$), this means that the complete list of values R$_1$ can choose is information that is encoded in $|\Psi_0\rangle$, and so this information is locally available at S.

   Within the simulation, each system will carry independent hidden variables locally, although it may carry many possible values in parallel.
   \begin{itemize}
   \item System r$_1$ carries all three random values from R$_1$ locally to M$_1$.
   \item System s$_1$ carries a list of amplitudes for each of the four possible outcomes of the measurement, given the four values of $q_1 \in \{|0\rangle,|1\rangle\}$ and $q_2 \in \{|0\rangle,|1\rangle\}$ in $|\psi\rangle$, and the three possible values of r$_1$, for a total of 48 different amplitudes.
   \item When r$_1$ and s$_1$ converge at M$_1$, there are twelve possible futures carried by m$_1$, and the quantum probabilities determine how the amplitudes for the three values carried by r$_1$ and the 48 carried by s$_1$ are matched up as the histories of these twelve.  Specifically, sixteen values carried by s$_1$ each become correctly correlated with their corresponding value carried by r$_1$, and the amplitudes from the four values of $q_1$ and $q_2$ interfere to form just four possible futures for each value of r$_1$.
   \item Likewise for r$_2$, s$_2$, and m$_2$.
   \item When m$_1$ and m$_2$ converge at E, there are 72 possible futures for E, and the quantum probabilities determine how the values from m$_1$ and m$_2$ become correlated as the histories of E.  Because of the entanglement correlation of $|\psi\rangle$, each value carried by m$_1$ has zero probability to be correlated with six of the values carried by m$_2$ (or vice versa), and thus only 72 (not 144) possible combinations of values from m$_1$ and m$_2$ remain as possible histories of E.
   \item It is noteworthy that while m$_1$ (m$_2$) carries twelve possible values, it also carries 48 different possible histories, and while E carries 72 possible values, it also carries 288 possible histories.
\end{itemize}

   Thus, by using advance knowledge of all possible interactions in a system's future light cone, it is possible for each system to locally carry information that determines its actual future - the hidden variable.  When systems interact locally in the future, the simulation simply matches the pre-corresponding possibilities carried by each system.  Using this locally causal and deterministic mechanism, the simulation is able to compute the list of all possible futures of the universe starting from $|\Psi_0\rangle$ alone.

\begin{figure}
\centering
\resizebox{\textwidth}{!}{
\begin{tabular}{c|lll}
  \hline$r_1$ & $|\mu_i\rangle^{r_1}_,$ & $\langle \phi_{ik}|^{m_1}_,$ & any \textbf{one} value of each of $i \in \{1,2,3\}$ and $k \in \{1,2,3,4\}$.  \\
   $r_2$ & $|\mu_j\rangle^{r_2}_,$ & $\langle \phi_{jl}|^{m_2}_,$ & any \textbf{one} value of each of $j \in \{1,2,3\}$ and $l \in \{1,2,3,4\}$. \\
   $s_1$ & $|\psi\rangle^{s_1}_,$ & $\langle \phi_{ik}|^{m_1}_,$ & the same $i$ and $k$ as above. \\
   $s_2$ & $|\psi\rangle^{s_2}_,$ & $\langle \phi_{jl}|^{m_2}_,$ & the same $j$ and $l$ as above. \\
   $m_1$ & $|\phi_{ik}\rangle^{m_1}_,$ & $\langle\Omega_{ijkl}|^E_,$ & the same $i,j,k$, and $l$ as above. \\
   $m_2$ & $|\phi_{jl}\rangle^{m_2}_,$ & $\langle\Omega_{ijkl}|^E_,$ & the same $i,j,k$, and $l$ as above. \\
   $E$   & $|\Omega_{ijkl}\rangle^E_,$ & $\langle ? |_,$ & the same $i,j,k$, and $l$ as above.  \\
\hline
\end{tabular}}
\caption{Information carried by different physical systems in the local realist model is shown.  Pre-selected states are shown as kets and post-selected states as bras.  The locally deterministic simulation only allows values of $i,j,k$, and $l$ that obey the entanglement correlations of $|\psi\rangle$, since others have zero amplitude.}
\label{Inflation}
\end{figure}

      Finally nature randomly selects one such future to be the one we call reality, according to the usual probability rules.  Once a specific selection is made, each of the signals r$_1$, r$_2$, s$_1$, s$_2$, m$_1$, and m$_2$ carries only a single value forward in time, as shown in Fig. \ref{Inflation}.  This is a standard feature of realism, but we can also make the case that they carry information about the outcome of a random quantum event in the future, which is to say the post-selected state, since this information was used locally in the simulation, and in this model, quantum mechanics is the manifest residual evidence of the simulation.

   Concerning weak values and weak measurements, this means that the weak value {\it is} defined using information about the future outcome of a random quantum event, and that weak measurements {\it may} truly probe this hidden information.  In this view, it is the unavoidable noisiness of weak measurements that protects the experienced randomness of quantum physics, and in turn the illusory experience of free will.

\section{Parallel Lives}

The Parallel Lives interpretation of quantum mechanics of Brassard and Raymond-Robichaud \cite{brassard2013can} is a many-worlds-type theory \cite{everett1973theory, vaidman_site} that ascribes ontological reality to all of the possibilities that are carried by r$_1$, r$_2$, s$_1$, s$_2$, m$_1$, and m$_2$. It is important to note that in this picture, the entire universe is not divided into many worlds, but rather the universe is comprised of many interacting quantum systems, and these can individually and independently experience parallel lives.  There is no wavefunction collapse or classical limit in this interpretation, but again a single universal wavefunction that contains a superposition state of all systems.   The model described here may not exactly duplicate what these authors intend, but the ideas are very similar, and they deserve attribution.

The guiding principle of this interpretation, as we explain it here, is that the experience of every life of every system is local and continuous, and that no life is experienced as a superposition state.  This also implies that each life can only experience product states, since entangled states require superposition.  When two systems carrying multiple lives interact, all possible pairings of those lives become viable interaction histories for the future lives of those systems, with the probability amplitude of each history determined by the usual quantum rules.  The correlations predicted by entanglement are thus obeyed when future interactions occur.  For each of the future lives individually, this is subjectively experienced as a random collapse onto one outcome - possibly with several interfering histories.

In this model, the different lives of a system are the orthogonal terms in a superposition state, which are completely oblivious of one another, and each life is weighted by its probability amplitude from standard quantum mechanics.  The weight here now acts as a `measure of existence,' meaning that some lives exist `more' than others \cite{vaidman2012probability}.  To conceptualize this, suppose there are truly a very large number of sub-lives, like the particles in a fluid, and that the fluid may flow more into one channel than another, meaning that a particle in the fluid can be more likely to find itself in one channel than another.  The lives we speak of here are the different channels, which correspond to the orthogonal terms of the superposition, and the amount of fluid in each channel corresponds to the probability amplitude associated with each term.  The `particles' in the fluid experience the channel, but are oblivious of one another.  The total amount of fluid is conserved.

This also implies an important refinement to the general many-worlds interpretation, which is that all quantum systems must have a preferred basis at any given time in order to define the superposition that dictates which parallel lives are being experienced for each system.  The evidence of this is in some sense anthropic, since the life that a given system experiences must be an eigenstate in the preferred basis \cite{vaidman1998schizophrenic}. There are also other compelling reasons to suppose that the universal wavefunction has a preferred basis at any given time \cite{schwindt2012nothing}.

Fig. \ref{Parallel} explicitly shows which lives of each system are propagating according to local causality and determinism in the Parallel Lives interpretation.

\begin{figure}
\centering
\resizebox{\textwidth}{!}{
\begin{tabular}{c|ll}
\hline
  $r_1$ & $|\mu_i\rangle^{r_1}_,$   & all three values of $i \in \{1,2,3\}$ (3 different lives of $r_1$). \\
   $r_2$ & $|\mu_j\rangle^{r_2}_,$  & all three values of $j \in \{1,2,3\}$ (3 different lives of $r_2$).\\
   $s_1$ & $|\psi_{q_1,q_2}\rangle^{s_1}_,$ & all values of $q_1 \in \{0,1\}$ and $q_2 \in \{0,1\}$ (4 lives of $s_1$).  \\
   $s_2$ & $|\psi_{q_3,q_4}\rangle^{s_2}_,$ & all values of $q_3 \in \{0,1\}$ and $q_4 \in \{0,1\}$ (4 lives of $s_2$).  \\
   $m_1$ & $|\phi_{ik}(|\mu_i\rangle^{r_1}_,|\psi\rangle^{s_1} )\rangle^{m_1}_,$ & all 12 combinations of $i$ with $k \in \{1,2,3,4\}$ (12 lives of $m_1$).\\
   $m_2$ & $|\phi_{jl}(|\mu_j\rangle^{r_2}_,|\psi\rangle^{s_2} )\rangle^{m_2}_,$ & all 12 combinations of $j $ with $l \in \{1,2,3,4\}$ (12 lives of $m_2$).\\
   $E$   & $|\Omega_{ijkl}(|\mu_i\rangle^{r_1}_,|\psi\rangle^{s_1}_,|\mu_j\rangle^{r_2}_,|\psi\rangle^{s_2})\rangle^E_,$ & all 72 combinations of $i,j,k$, and $l$ consistent with the entanglement  \\
   & & \ \ \ correlations of $|\psi\rangle$. (72 different lives of $E$) \\
   \hline
\end{tabular}}
\caption{Information carried by different physical systems in the Parallel Lives model.  Each system propagates locally carrying all possible lives and a record of past local interactions, as shown in parentheses.  At $E$, the past lives of m$_1$ and m$_2$ become correlated with one another with amplitude given by the usual quantum probability law in order to determine the future lives of $E$ - and the amplitude is zero that the entanglement correlations will be violated.}
\label{Parallel}
\end{figure}

   This interpretation has given up conventional realism in lieu of the ontological realism of many systems experiencing many parallel lives.  It makes no use of cosmic inflation, but it is locally causal and deterministic, as we now explain.

   Considering Fig. \ref{Causality}, in the Parallel Lives model, each qubit of $|\psi\rangle$ has lives for $|0\rangle$ and $|1\rangle$, meaning four lives are carried by each of the systems s$_1$ and s$_2$.  It is important to note that each life in each system carries a record of its local interaction history, which will allow the correct entanglement correlations to be obeyed locally when later interactions between systems occur.

   R$_1$ and R$_2$ each create three possible lives that are carried locally by r$_1$ and r$_2$.

   When the s$_1$ and r$_1$ converge at M$_1$, all of the possible lives in each system become correlated according to the standard quantum probability rules (tracing qubits 3 and 4 out of the $|\psi\rangle$), which in this case means that all twelve combinations occur with equal amplitude.  Each of the twelve combinations leads to a complex amplitude for the twelve outcomes of the PMS measurement at M$_1$ (four possible outcomes for each of three possible measurements), and these sum (interfere) to determine the actual amplitude of each of the twelve lives carried by m$_1$.  It is important to note here that each life of m$_1$ includes a combination of four possible histories from s$_1$, and so there are now truly 48 distinct histories for the 12 lives in m$_1$ (depending on what basis was used for $q_1$ and $q_2$, some of these histories may have zero amplitude, but the 12 lives are independent of this detail).  Likewise for s$_2$, r$_2$, and m$_2$, and there is no special correlation between the lives of the systems in the two arms of the experiment, which we assured by tracing out the space-like separated qubits.

   When m$_1$ and m$_2$ converge at E, the entanglement correlations of $|\psi\rangle$ now dictate that the amplitude for lives of E with $q_1 \neq q_3$ or $q_2 \neq q_4$ in their history is zero.  Nevertheless, every life from one arm finds the correct proportion of lives from the other arm to become correlated with, so there are no `loose ends.'  This correlation of past histories of different interacting systems produces the expected entanglement correlations in the outcomes observed by each life of E.  This means that for each of the twelve lives carried by m$_1$, only six of the lives carried by m$_2$ can become correlated as the history of a life of E, and vice versa, so there are 72 total lives carried by E (rather than 144).  Each of these lives again includes a combination of four different correlated histories from s$_1$ and s$_2$, meaning there are 288 different histories for E.

   Let us recapitulate the experience of following a single life along a single local causal path of several systems.  Let us begin with the life $|0,0\rangle^{s_1}$ carried by s$_1$, which is a product state of two qubits.  When this system arrives at M$_1$, it becomes correlated to choice $|\mu_1\rangle^{r_1}$, and outcome $|\phi_{11}\rangle^{m_1}$ is obtained.  Finally, when this system m$_1$ arrives at E, it becomes correlated to $|\phi_{11}\rangle^{m_2}$, which obeys the entanglement correlation condition that $ZI = +1$.  Thus, following a single life as it propagates by local causality, we see the experience of collapse events when one of several possibilities is obtained, and also that entanglement correlations are obeyed locally as part of this experience.

   In summary, the locally deterministic propagation of systems carrying multiple lives, along with tracking of the interaction-histories of each life, allows quantum probabilities to be computed locally, and explains the quantum correlations experienced by each of the 72 lives of E.  These correlations thus violate the Bell inequalities while maintaining local causality and determinism.

   The Appendix contains more information about the Parallel Lives model, and goes through several important examples in detail.

   In this picture multiple lives can be carried by a given system, but they always carry this information locally forward in time.  This means that a weak measurement is only biasing the experienced result of a future quantum random event, but it cannot probe the predetermined outcome of the event, because this information is not present in the system (as it was in the simulation of the inflation model).

   \section{Parallel Lives with Cosmic Inflation}

Finally, let us consider the union of the inflation model with the parallel lives model. Now, rather than being forced to simulate all possible future realities in order to choose one at the time of the singularity, we propose that the simulation itself is ontologically real, and all possible futures exist in parallel.  The locally deterministic mechanism we described above governs the evolution of all such futures, which is to say the parallel lives of all systems in the universe.  The only fundamental difference here is that all systems and lives have the initial state of the universe in their interaction-history, which, as we have seen with the Parallel Lives interpretation, means $|\Psi_0\rangle$ can govern the amplitudes for different lives to become correlated anywhere in the future, according to the usual rules of quantum probability.

\begin{figure}
\centering
\resizebox{\textwidth}{!}{
\begin{tabular}{c|lll}
\hline
  $r_1$ & $|\mu_i(|\Psi_0\rangle)\rangle^{r_1}_,$   & $\langle \phi_{ik}|^{m_1}_,$ & all three values of $i \in \{1,2,3\}$ (3 different lives of $r_1$). \\
   $r_2$ & $|\mu_j(|\Psi_0\rangle)\rangle^{r_2}_,$  &$\langle \phi_{jl}|^{m_2}_,$ &  all three values of $j \in \{1,2,3\}$ (3 different lives of $r_2$).\\
   $s_1$ & $|\psi_{q_1,q_2}(|\Psi_0\rangle)\rangle^{s_1}_,$ & $\langle \phi_{ik}|^{m_1}_,$ & all 48 combinations of $i$ with $k \in \{1,2,3,4\}$,  \\
    &  & & \ \ \ $q_1 \in \{0,1\}$, and $q_2 \in \{0,1\}$ (48 lives).  \\
   $s_2$ & $|\psi_{q_3,q_4}(|\Psi_0\rangle)\rangle^{s_2}_,$ & $\langle \phi_{jl}|^{m_2}_,$ & all 48 combinations of $j $ with $l \in \{1,2,3,4\}$ \\
   &  & & \ \ \ $q_3 \in \{0,1\}$, and $q_4 \in \{0,1\}$ (48 lives).\\
   $m_1$ & $|\phi_{ik}(|\mu_i\rangle^{r_1}_,|\psi\rangle^{s_1}_,|\Psi_0\rangle )\rangle^{m_1}_,$ & $\langle\Omega_{ijkl}|^E_,$ &  all 12 combinations of $i$ with $k \in \{1,2,3,4\}$ \\
   &  & & \ \ \ (12 different lives of $m_1$).\\
   $m_2$ & $|\phi_{jl}(|\mu_j\rangle^{r_2}_,|\psi\rangle^{s_2}_,|\Psi_0\rangle )\rangle^{m_2}_,$ & $\langle\Omega_{ijkl}|^E_,$ &  all 12 combinations of $j $ with $l \in \{1,2,3,4\}$ \\
   &  & & \ \ \ (12 different lives of $m_2$).\\
   $E$   & $|\Omega_{ijkl}(|\mu_i\rangle^{r_1}_,|\psi\rangle^{s_1}_,|\mu_j\rangle^{r_2}_,|\psi\rangle^{s_2}_,|\Psi_0\rangle)\rangle^E_,$  & $\langle ? |_,$ & all 72 combinations of $i,j,k$, and $l$ consistent with the \\
   &  & & \ \ \ entanglement correlations of $|\psi\rangle$ (72 different lives of $E$). \\
   \hline
\end{tabular}}
\caption{Information carried by different physical systems in the combined Parallel Lives and cosmic inflation model is shown.  Pre-selected states are shown as kets and post-selected states as bras.  Each system propagates locally carrying all possible lives and a record of past local interactions, as shown in parentheses.  At M$_1$, the past lives of r$_1$ and s$_1$ become correlated with one another according to the usual quantum probability law in order to determine the future lives of m$_1$, and likewise for M$_2$.  And again at $E$, the past lives of m$_1$ and m$_2$ become correlated with one another according to the usual quantum probability law in order to determine the future lives of $E$ - and the probability is zero that the entanglement correlations will be violated.}
\label{ParallelInflation}
\end{figure}

   In this version of the parallel lives model, information about post-selected states is always available, and weak measurements {\it may} probe information about the future of a quantum random event.  Importantly, a system divides into parallel lives prior to the interaction where the post-selection is actually realized.  The system is still prevented from learning about this advance splitting by the noisiness of weak measurements.  In particular, s$_1$ and s$_2$ carry 48 lives each (as they did in the inflation simulation) instead of four lives as in the Parallel Lives model, as shown in Fig. \ref{ParallelInflation}.

   Generally speaking, all systems are split into all possible future realities starting at the time of the singularity, which is how the simulation would be able to choose one at that time in the superdeterministic model above.

   This unified picture provides a locally causal and deterministic model of quantum mechanics that violates traditional realism by supposing an ontology of many systems carrying many parallel lives.  The fundamental difference with the usual parallel lives model is that in this model, hidden variables predicting all possible (parallel) outcomes do exist and are carried by physical systems obeying local causality.

   Even though this is a deterministic model which cannot truly support free will, we argue that the experience of weighted randomness is philosophically indistinguishable from the experience of free will - and this is the experience of each individual life of a quantum system.  In this way, free will emerges from our ignorance of which among a sea of possible futures will be the one we experience - and again this ignorance is guaranteed by the unavoidable noisiness of weak measurements.

   The same argument concerning free will applies to the usual parallel lives interpretation as well, except that there is no requirement that weak measurements be noisy, since there is no post-selected state that must be obscured.

   \section{Discussion and Conclusion}

   All of the models we have discussed here have many features in common - they are all deterministic models with local causality, and all lack a classical limit or any notion of wavefunction collapse.  Nevertheless, the experience of randomness and of collapse are completely explained in each case, and decoherence explains the experience of the classical world for macroscopic objects due to dissipative interactions with the environment.

      The assumption of a singularity in the past light cone of all events allows hidden variables predicting all outcome to be carried locally by all systems, and perhaps weakly measured.  Without this assumption, there is no need for the hidden variables to exist at all - the evolution mechanism is locally causal and deterministic in either case.

   Some researchers are critical of many-worlds-type models because the proliferation of parallel worlds or lives that we do not see seems to violate some notions of Occam's Razor.  But then the ability of models (or simulations) like this to provide locally causal and deterministic explanations of quantum randomness, collapse, and Bell nonlocality, without resorting to some new axioms regarding a classical limit or No-Signalling, can be argued to be favorable for just the same reason.  In particular, the classical limit is playing a trick analogous to parallel lives, by explaining how there is a quantum world, but one that we never directly see.  Why introduce a classical limit, or an explicitly nonlocal interaction mechanism when both of these can be avoided \cite{deutsch2012vindication}?

   Consider a version of the Wigner's friend thought experiment, in which Schr\"{o}dinger is inside a sealed room that also contains a sealed box with a cat and a vial of poison.  At a certain pre-arranged time $t_1$, the cat will either be killed by the poison or saved from it by the outcome of a microscopic random quantum event, at a subsequent time $t_2$ Schr\"{o}dinger will open the box and learn the fate of the cat, and finally at $t_3$ Wigner will open the door to the room and learn what has happened inside.  What can we say about the continuous experiences of the three macroscopic entities in this story?

   If there is a classical limit for macroscopic objects, then the cat was either alive or dead, but was never in a superposition state, and the same is true of Schr\"{o}dinger within the room - collapse only needed to occur at $t_1$.  In the superdeterministic model we presented above, all of the outcomes were again predetermined, and so there was just a single outcome and a single experience, but no collapse occurs during the experiment.

   Suppose we relax the classical limit, and allow the cat to enter a superposition state at $t_1$, and then Schr\"{o}dinger and the cat enter an entangled superposition state at $t_2$, but still Wigner causes a collapse at $t_3$.  What was Schr\"{o}dinger's experience of being in a superposition state between $t_2$ and $t_3$?  Was this a sort of dreamlike state in which both realities were somehow simultaneously true?  What would it feel like \cite{vaidman1998schizophrenic}?  Or if there were two parallel versions of Schr\"{o}dinger, each experiencing one well-defined outcome, what happens to the other one when the collapse occurs?  Is it simply snuffed out of existence?  Its proponents may not agree, but this seems to be roughly the story from the Consistent Histories interpretation of quantum mechanics \cite{griffiths1984consistent}.

   However, if there is no collapse at all, as in the parallel lives models discussed here, then the two parallel versions of Schr\"{o}dinger and the cat both exist, and Wigner simply joins the entangled superposition state at $t_3$.  Neither version of Schr\"{o}dinger and the cat is terminated.  Both parallel versions simply continue to exist, oblivious of one another forevermore.

   These parallel lives models can be seen as a type of local fluid dynamics model, reminiscent of Madelung \cite{madelung1926quantum}, where the fluid carries probability amplitudes and parallel lives, and obeys a continuity relation.  The streamlines of this model will be like Bohmian trajectories (see also \cite{bostrom2014quantum}), all representing parallel lives of the system, but without any need of a nonlocal interaction.  It also seems likely that this will be the quasi-continuum limit, or fluid-like limit, of the Many-Interacting-Worlds-type models of Poirier et al \cite{schiff2012communication} and Wiseman et al \cite{hall2014quantum}, which is to say that they will reduce to local fluid-like dynamical models, in which the parallel lives mechanism described above resolves any Bell-type nonlocality involving multiple entangled systems.  Perhaps the discrepancy between locality in configuration-space and locality in position-space within these models will vanish once it is understood how gravity and/or inflation fit into the story.  In the end, the interaction between the worlds may simply be local interference of waves in the fluid - the Schr\"{o}dinger equation is after all a type of local dynamical wave equation.

This picture attempts to unify key ideas from several interpretation of quantum mechanics.  The notion of a conserved probability fluid with local dynamics, and no need for collapse or a classical limit, has its own elegant simplicity when viewed through the lens of Occam's Razor, even if we observers do not directly experience all facets of this ontology.

   While this article has focused on a particular class of interpretations that we find appealing, we are not true believers of any one model, as we do not see sufficient experimental evidence to support one over another.  We may be guided by ideals like elegance or beauty as we develop and pursue new models, but this is only the artistic inspiration of scientific research and should not be conflated with rigorous argumentation.

   There are a number of seemingly self-consistent interpretations of quantum mechanics - none significantly more or less correct than the others, given experimental data that is presently available.  It is the view of this author that any of these may lead to new and unique insights about quantum physics, which may in turn lead to new experiments, and new data that could allow us to narrow down the possibilities.  One might say that the current state of the art is a superposition of many interpretations, and insofar as they can be useful tools for scientific research, any self-consistent theory has value.

   \textbf{Acknowledgments:}  We would like to thank Yakir Aharonov, Jeff Tollaksen, and Justin Dressel for interesting conversations that impacted this work.  This research was supported (in part) by the Fetzer Franklin Fund of the John E. Fetzer Memorial Trust.

  \bibliographystyle{ieeetr}
%\bibliography{Citation_Database}
\bibliography{Inflation_ParallelLives.bbl}

   \section{Appendix}
   Here we go through several examples intended to clarify the parallel lives model.  We discuss entangled states, interference effects, memory and quantum erasers, quantum relativity, and the possibility of non-unitary evolution.

\subsection{Entangled States}

   Here we explain the detailed behavior of a simple entangled state in the parallel lives model.  Consider the entangled state,
   \begin{equation}
   |\psi\rangle = \frac{4}{5} |0\rangle |0\rangle + \frac{3}{5}|1\rangle |1\rangle.
   \end{equation}
   The two qubits are sent to space-like separated detectors A and B and measured in the computational basis, and then the results are subsequently collected by the experimenter and compared.  There are several important details to take note of:
   \begin{itemize}
   \item There is a preferred basis for the two qubits, and it may not be the computational basis in which the state is represented above.  For simplicity in this example, suppose it is the computational basis.
   \item This means that each qubit is experiencing two different lives, $|0\rangle$ and $|1\rangle$, and 16/25 of the total probability fluid is in the channel for $|0\rangle$ and 9/25 of it is in the channel for $|1\rangle$ - although the two lives are oblivious to the amount of fluid that flows along with them, and oblivious to one another.  Each system carries a record of $|\psi\rangle$ in its local interaction history.
       \item  When the measurement is performed, detector A reports $|0\rangle$ with probability 16/25, and $|1\rangle$ with probability 9/25, as it becomes correlated with one of the two lives of that qubit.  The same is true for the qubit at detector B.  These are just the usual probabilities predicted by quantum mechanics for the reduced density matrix of each qubit (i.e... oblivious of what happens to the other qubit).
           \item The physical system that carries the result of measurement A back to the experimenter is now experiencing two different lives, with the same relative proportion of probability fluid, and likewise for the system that carries the result of measurement B.
               \item Because of the record of $|\psi\rangle$ carried by both systems, the life of system A in which the result of measurement A was $|0\rangle$ only becomes correlated with the life of system B with the result $|0\rangle$, and likewise for the lives with $|1\rangle$.  Importantly, the amount of fluid is identical in the channels that become correlated, and now the experimenter has two lives, one in which the result $|0\rangle |0\rangle$ was obtained (with 16/25 of the fluid), and one in which the result $|1\rangle |1\rangle$ was obtained (with 9/25 of the fluid).
                   \item  Clearly both lives of the experimenter observe the correct entanglement correlation for $|\psi\rangle$, and the total amount of probability fluid is conserved, which is to say that all lives of all systems are continuous.  The entire mechanism is locally causal and deterministic.
   \end{itemize}

  It is also worthwhile to consider the less simple case in which the preferred basis is the $\sigma_x$ eigenbasis, $|+\rangle = \frac{1}{\sqrt{2}} |0\rangle + \frac{1}{\sqrt{2}}|1\rangle$, and $|-\rangle = \frac{1}{\sqrt{2}} |0\rangle - \frac{1}{\sqrt{2}}|1\rangle$.  Then we can rewrite $|\psi\rangle$ as,
  \begin{equation}
   |\psi\rangle = \frac{7}{10} |+\rangle |+\rangle +  \frac{1}{10} |+\rangle |-\rangle + \frac{1}{10}|-\rangle |+\rangle + \frac{7}{10}|-\rangle |-\rangle.
   \end{equation}
   In this basis, each of the two lives of each qubit carries half of the probability fluid, which we see by summing the mod-squared amplitudes of the terms with each single-qubit state $|+\rangle$ or $|-\rangle$,
\begin{equation}
\left(\frac{7}{10}\right)^2 + \left(\frac{1}{10}\right) ^2 = \frac{1}{2}.
\end{equation}
    Again each system carries a record of the state $|\psi\rangle$ in its local interaction history (i.e. a record of what happened when the two systems did interact locally in the past).

   When the measurements A and B occur, the probability fluid now mixes (interferes) as it flows from the channels of the $\sigma_x$ basis into the channels of the computational ($\sigma_z$) basis of the measurement, and the record of $|\psi\rangle$, written in the computational basis, determines how much fluid flows into each channel, so that again 16/25 flows in the channel for $|0\rangle$ and 9/25 in the channel for $|1\rangle$.  Notice that regardless of the preferred basis for the qubits of the original Bell state, after the measurement, the preferred basis is the computational basis.

   This case also raises an interesting question.  Each system carries 1/2 $|+\rangle$ and 1/2 $|-\rangle$, but then interference at the measurement causes 16/25 to flow into $|0\rangle$ and 9/25 to flow into $|1\rangle$, but what was the probability for a $|+\rangle$ life to find itself in the $|0\rangle$ channel?  This detail does not seem to be specified by the current model - it is clear that all of the fluid is conserved, but due to the interference, it is not clear how the two pre-measurement channels flow into the two post-measurement channels. Perhaps a Bohmian-like treatment of the individual trajectories can answer this question.

   In the above example, the order of events played no important role, and this description works the same is any relativistic reference frame.  All physical systems propagate at typical luminal or subluminal speeds and carry with them all of the information that is needed to correctly correlate the parallel lives when multiple systems interact locally.

   \subsection{Quantum Relativity}

   In order to conceptualize the universe of many parallel lives, we also introduce the notion of quantum relativity, which formalizes our earlier discussion of the Wigner's friend thought experiment.  Recall that in this model, there is just one universal wavefunction that describes an elaborate superposition state for all systems.  The experience of a life within the universal wavefunction is subjective - it is a special projection onto a single eigenstate of a given system in that system's preferred basis, or one specific channel through which probability fluid flows.  The complete life experience of a given life of a given system is exactly the complete record of its local interaction histories with other systems, and systems they had previously locally interacted with, etc..., and this experience is unique for every life of every system.

   For each life of each interacting system, the experience of an interaction is a collapse onto a product state of the interacting systems. However, because each system experiences splitting into parallel lives independently, whenever two systems interact locally, they can enter a superposition relative to remote systems.  Specifically, when systems A and B interact locally, the individual lives of A and B experience a collapse into a product state of the two systems, but relative to system C which did not participate in the interaction, systems A and B have not collapsed, but rather undergone a unitary transformation into an entangled superposition state whose terms are the different product-state lives the two systems are experiencing.  This product basis is also the new preferred basis for both systems.  If system C interacts with system A or system B in the future, the individual lives of C will also experience a collapse, but relative to system D there is no collapse, this is now just an entangled superposition of all three systems A, B, and C, and so on...

   Within a single universal wavefunction, quantum relativity is the fact that an interaction that is experienced as a collapse by the lives of the interacting systems, is instead experienced as a unitary evolution by all other systems in the wavefunction, and both experiences are perfectly consistent.  At the level of the universal wavefunction, there is never a collapse - which conventionally leads us to conclude that only unitary evolution occurs.

   \subsection{The Mach-Zender Interferometer, Wheeler's Delayed Choice, and Elitzur-Vaidman Interaction-Free Measurements}

Consider a Mach-Zender interferometer, tuned for perfect constructive and destructive interference respectively.  After passing through the first beam splitter, half of the fluid flows down each arm of the beam splitter - the two arms are orthogonal in position and momentum space.  At the second beam splitter there is interference between the fluid coming from the two arms.  Interference occurs because the fluid coming from each arm is once again in the same place, with the same momentum, and thus no longer orthogonal - interference occurs only when non-orthogonal fluid flows together.

This probability fluid picture, with half of the fluid flowing through each arm of the interferometer, also provides a clear explanation for Wheeler's Delayed Choice experiment \cite{wheeler1978past,jacques2007experimental}, and the Elitzur-Vaidman interaction-free measurements \cite{elitzur1993quantum}, as locally causal and deterministic.

In a Wheeler delayed-choice experiment using a Mach-Zender, if the second beam splitter is not present, the 1/2 measure of fluid from the two arms simply remains orthogonal and crosses paths without any interaction, arriving at both detectors.

In interaction free measurement, the fluid from one arm of the interferometer is simply blocked by the `bomb,' which means it cannot participate in the interference at the second beam splitter.  For the lives that experience the interaction free measurement, it is exactly because this probability fluid fails to arrive that the presence of the bomb is revealed.  On the other hand, the lives that were blocked most certainly {\it did} experience interacting with the `bomb.'

\subsection{Interference and Memory, Quantum Erasers}

   Now, if a system is capable of encoding a memory into Hilbert space of which arm it took through the beam splitter, then this information makes the fluid that recombines at the second beam splitter orthogonal, and thus there is no interference.  The Hilbert space into which the path-information is encoded can be internal to the system traversing the interferometer, or it can be encoded into the environment by an interaction - which is the explanation due to decoherence.

  A general quantum eraser experiment \cite{scully1982quantum, walborn2002double} involves an entangled pair of systems, which we call {\it signal} and {\it idler}, the orthogonal information is encoded into the idler system and the signal system interferes with itself (through an interferometer, diffraction grating, etc...).  If the orthogonal information is measured in the idler system, no interference is observed in the signal system.  On the other hand, if the orthogonal information is `erased' by a unitary operation before the idler is measured, then the interference pattern is observed in the signal.

  In the delayed choice quantum eraser \cite{kim2000delayed}, the decision to erase the information is postponed until well after the signal photon has been recorded. The post-selected signal data corresponding to the cases in which the idler was erased reveal complementary interference patterns, while the data corresponding to cases where the idler was not erased show no interference at all.

  This is perfectly consistent when one considers how the parallel lives of entangled systems become correlated as the experimenter collects all of the data.  Nevertheless, it is a subtle and nontrivial case that warrants a detailed description here.  We will consider the Delayed Choice Quantum Eraser experimental setup of Kim et al, in which a photon passes through a double slit and is then incident on a BBO crystal beyond either slit.  At each site, spontaneous parametric down conversion converts an incoming photon into an anticorrelated entangled pair of photons, and we define a signal and idler photon for each site.  The two signal paths are directed to a screen where they can interfere, and are then measured.  The two idler paths take a longer route, such that they are still in mid-flight when the signal photon is measured.  After this, the idler paths pass through beam splitters, which effectively `decide' whether or not the orthogonal information about which slit the photon passed through is erased.  In one case the two idler paths interfere at a third beam splitter which erases the orthogonal information, while in the other the idler from each path is directly measured, revealing exactly which path the photon took.  The coincident arrival of the signal and idler photons are tracked so that the signal data can later be post-selected into separate sets for each of the four possible detectors the idler arrived at.  In the two cases where the orthogonal which-slit information was not erased, the signal data shows no interference pattern.  In the two cases where the orthogonal information was erased, the signal data reveals two different interference patterns, complementary in sense that their fringes are $\pi$ out of phase.  Because of this complementarity, the interference pattern is completely hidden in the full data set.

  Now we discuss the probability fluid picture of this experiment.  The fluid for the incident photon arrives at the first beam splitter and divides in half.  The 1/2 fluid at each BBO site creates a signal and idler photon which each carry a 1/2 measure of probability fluid.  The two signal channels interfere and are measured - all signal fluid arrives at the screen.  The 1/2 measure of idler fluid from site A arrives at beam splitter A and divides into 1/4 going to detector A and 1/4 going to beam splitter C.  Likewise, the 1/2 measure of idler fluid from site B arrives at beam splitter B and divides into 1/4 going to detector B and 1/4 going to beam splitter C.  The idler fluid from each site arrives at beam splitter C and divides into 1/8 each, which interfere to become 1/4 going to each of the two `erased' detectors C$_1$ and C$_2$.  Given our earlier discussion of entanglement correlations in the parallel lives model, we can see that when the signal and idler data are collected, the experimenter now experiences many parallel lives in which the entanglement correlations are obeyed, meaning that pattern of erased idler photons matched with signal photons will reveal interference, and the pattern of unerased idler photons matched with signal photons will show no interference.  It is important to emphasize that every life of each system (signal and idler) exists independently before the experimenter collects the data, and quantum theory tells us the probability (including entanglement correlations) that each possible pairing of the lives of the two systems will be experienced by the experimenter, and furthermore that every signal life finds a match with an idler life.

  We should emphasize the difference in this model between the Quantum Eraser and the Wheeler's delayed choice, which seem more closely related than they are.  In the delayed choice experiment, the presence or absence of the second beam splitter changes the quantity of fluid that flows to each detector - specifically, if the beam splitter is present, no fluid at all flows to one of the detectors due to destructive interference.  After the photon passes through the first beam splitter, adding the second beam splitter causes interference which effectively erases the `which-path' information, while leaving out the second beam splitter allows the `which-path' information to be directly measured.

  Because the delayed-choice quantum eraser experiments use entangled states to encode and potentially erase the orthogonal information, the same quantity of probability fluid always flows through the parts of the experiment, and no interference patterns are visible at this level.  The interference patterns only emerge when the lives of the signal and idler become correlated according the the usual quantum probabilities.

   As an aside, it is worth mentioning that in all discussions of the Mach-Zender Interferometer, or the double-slit experiment, we have made an implicit assumption that when the particle is incident on a beam splitter (double-slit), the particle enters a superposition that is {\it not} entangled with the beam splitter (double slit).  If it were entangled in this way, this would mean the device itself had encoded orthogonal information, and it would never be possible to see an interference pattern.  Indeed, double-slit experiments have been performed in which atoms near the slits can acquire a record of which slit the particle passed through, and the visibility of the interference fringes is measurably reduced.  Similar loss of visibility is observed in the the large-molecule double-slit experiments of Arndt et al \cite{arndt1999wave, brezger2002matter,nairz2003quantum, hackermuller2004decoherence}, as various decoherence issues become increasingly difficult to avoid.  Interactions can create a superposition for one system, or an entangled superposition of several systems.  However, in the entangled case, if the system has decohered and interacted with the environment, then the interference vanishes because each life of the experimenter sees only one term in the superposition of the experimental system.

   \subsection{Irreversible Processes and Non-Unitary Evolution}

      In this picture, the unitary evolution seems to follow from conservation of probability fluid, but fluid conservation alone is actually a significantly weaker constraint.  For example, consider an irreversible logical process like the erasure of an isolated single-qubit system, as an interaction in which all fluid from two distinct channels flows together into a single channel.  The fluid is still conserved in this process, and indeed the trajectories of individual particles in the fluid are reversible, but the experience of a particle in the fluid would be oblivious to which channel it had come from.  Internally, the particles carry a record of their local interaction history, but isolated single-qubit systems cannot encode any memory about which of the two channels they came from, and thus the present experience of a life is fully defined by the channel in which the fluid now flows - at the level of experience, the information about the past has been erased.  There is nothing about local causality, determinism, or our parallel lives guiding principle that would prevent this situation from occurring, but this operation is not, in general, represented by a unitary operator in Hilbert space - it is instead a projection operator.  This may allow for a parallel lives model that does not strictly adhere to the Church of the Larger Hilbert space view, in which there can be only unitary operations.

It is interesting that this could only occur when non-orthogonal fluid flows together, meaning the condition for interference is apparently also the condition for the projective interactions discussed above.

The black hole information paradox may also be avoided if we only impose the condition that probability fluid is conserved as it crosses the event horizon, rather than insisting on unitary evolution.

\end{document}